\documentclass[11pt,a4paper]{article}
\usepackage{authblk}
\usepackage[utf8]{inputenc}
\usepackage[pdftex]{graphicx}
\usepackage{graphicx,ae,aecompl}
\usepackage{amsmath,amsthm,amsfonts}
\usepackage{lscape}   
\usepackage{rotating}
\usepackage{rotfloat}
\usepackage{subfigure}
\usepackage{multirow}
\usepackage{booktabs}
\usepackage{ctable}
\usepackage{threeparttable}
\usepackage{booktabs}
\usepackage{makeidx}
\usepackage{framed}
\usepackage{color}
\usepackage[colorlinks=true,linkcolor=blue,allcolors=blue]{hyperref}

\title{\Large \bf A class of priors to perform asymmetric Bayesian wavelet shrinkage}

\author{Alex Rodrigo dos Santos Sousa}

\affil{Universidade Estadual de Campinas (UNICAMP)\\ Departamento de Estat\'istica, Brazil \thanks{ Sousa (asousa@unicamp.br)}}

\date{}

\begin{document}

\maketitle

\begin{abstract}
This paper proposes a class of asymmetric priors to perform Bayesian wavelet shrinkage in the standard nonparametric regression model with Gaussian error. The priors are composed by mixtures of a point mass function at zero and one of the following distributions: asymmetric beta, Kumaraswamy, asymmetric triangular or skew normal. Statistical properties of the associated shrinkage rules such as squared bias, variance and risks are obtained numerically and discussed. Monte Carlo simulation studies are described to evaluate the performances of the rules against standard techniques. An application of the asymmetric rules to a stock market index time series is also illustrated. \\

\noindent{\bf Keywords:} Wavelets, wavelet shrinkage, asymmetric priors, nonparametric regression, curve estimation. \\
%\noindent{\bf JEL classifications:} C22, C58, G15.\\

\end{abstract}

\newpage

\section{Introduction}
The application of wavelet-based methods in nonparametric regression models occurs in various real problems in chemometrics, electrical engineering, neuroscience and economics, among others. In all those areas, nonparametric regression arises when there is an unknown function of interest, for example the absorbance curve of a given substance and noisy points of this curve are observed. The goal is to estimate this underlying function. In this context, the standard procedure is to represent this unknown function in terms of some functional basis in such way that the curve estimation problem becomes a problem of estimating the finite number of coefficients of the representation. Wavelets are functions that satisfy some properties (for example, they integrate zero) and their dilations and translations compose a basis for the space of squared integrable functions. Some attractive features of the representation of a function in terms of a wavelet basis are that wavelets are well localized in both time and frequency domains, i.e, the significant coefficients are associated with important features of the function such as peaks, discontinuities and oscillations, and the representation is typically sparse, i.e, most of the wavelet coefficients are zero or close to zero. See Daubechies (1992) for a theoretical development of wavelets and their mathematical properties. Also consult Vidakovic (1999) for applications of wavelets in statistical modeling. 

In nonparametric regression models, one observes a sample of noisy points of an unknown function. When moving the original data to the wavelet domain by the application of a discrete wavelet transform (DWT), one obtains noisy versions of the wavelet coefficients of the representation, which are called empirical wavelet coefficients. Then a shrinkage or threshold rule is applied to the empirical coefficients in order to estimate the wavelet coefficients. This rule acts by reducing the magnitudes of the empirical versions, in the sense that small empirical coefficients occur due to the presence of noise in the null coefficients. Several wavelet shrinkage or thresholding rules are available in the literature, see Donoho and Johnstone (1994, 1995) and Donoho (1993, 1995) for the seminal works about wavelet shrinkage and the proposition of the so called soft and hard thresholding rules, as well as Vidakovic (1999) and Nason (2008) for descriptions of some the most commonly-applied shrinkage and thresholding rules. Although the available shrinkage and thresholding rules have been successful when applied to many real data applications and have attractive statistical and mathematical properties, they were developed under the supposition that the random errors in the original data are symmetrically distributed with zero mean. This paper deals with wavelet shrinkage under asymmetrically random noise, which can occur in practice.

We propose Bayesian wavelet shrinkage rules involving the mixture of a point mass function at zero and an asymmetric density function as prior to the wavelet coefficients, where this asymmetric density belongs to a class of densities composed of the asymmetric beta, Kumaraswamy, triangular and skew-normal densities, as proposed by Sousa (2022) regarding the asymmetric shrinkage rule under beta priors. In fact these four distributions were chosen to compose the class because they have interesting and complementary characteristics that allow flexibility in their associated shrinkage rules. For instance, their hyperparameters are associated with the amount of shrinkage to be imposed by the rule on the empirical coefficients, but in different ways according to the specific density, with bounded support (beta, Kumaraswamy and triangular) or support in the entire real set (skew normal). Thus it is possible to incorporate prior information regarding the sparsity and support of the wavelet coefficients in the shrinkage process when considering the proposed class of densities. 

This paper is organized as follows: the statistical model and the general development of the wavelet shrinkage process are described in Section 2. The proposed Bayesian shrinkage rules under asymmetric priors and their statistical properties are provided in Section 3. Simulation studies to evaluate the performance of the shrinkage rules are discussed in Section 4. An illustration of the proposed procedure in to a real dataset involving the São Paulo stock market index (IBOVESPA, in Portuguese) is shown in Section 5. The paper ends with final considerations in Section 6.

\section{Statistical models and wavelet shrinkage procedure}
Consider $n = 2^J$ ($J \in \mathbb{N}$) observations $(x_1,y_1),\cdots,(x_n,y_n)$ from the classical univariate nonparametric regression model
\begin{equation}\label{tmodel}
y_i = f(x_i) + e_i,
\end{equation}
where $x_i$ are scalars, $f$ is an unknown squared integrable function and $e_i$ are independent and identically distributed (iid) normal random errors with $\mathbb{E}(e_i)=0$ and $ \mathrm{Var}(e_i) = \sigma^2$, $\sigma > 0$. Thus $y_i$ are noisy observations from the unknown function $f$ at the locations $x_i$ and the goal is to estimate this function without assumptions about its functional structure. In vector notation, we can rewrite \eqref{tmodel} as
\begin{equation}\label{tmodel2}
\boldsymbol{y} = \boldsymbol{f} + \boldsymbol{e},
\end{equation}
where $\boldsymbol{y} = [y_1,\cdots, y_n]'$, $\boldsymbol{f} = [f(x_1),\cdots, f(x_n)]'$ and $\boldsymbol{e} = [e_1,\cdots, e_n]'$.

The standard procedure in wavelet shrinkage is to apply a discrete wavelet transform (DWT) to the original dataset in order to work in the wavelet domain. A DWT can be represented by an orthogonal transformation matrix $\boldsymbol{W}$ of dimension $n \times n$ that is applied on both sides of \eqref{tmodel2}. Since the DWT is linear, we obtain the following model in the wavelet domain
\begin{equation}\label{wmodel}
\boldsymbol{d} = \boldsymbol{\theta} + \boldsymbol{\varepsilon},
\end{equation}
where $\boldsymbol{d} = \boldsymbol{W}\boldsymbol{y} = [d_1,\cdots, d_n]'$ is the vector of the empirical wavelet coefficients, $\boldsymbol{\theta}= \boldsymbol{W}\boldsymbol{f} = [\theta_1,\cdots, \theta_n]'$ is the vector of the unknown wavelet coefficients and $\boldsymbol{\varepsilon} = \boldsymbol{W}\boldsymbol{e}= [\varepsilon_1,\cdots, \varepsilon_n]' $ is the vector of the random errors. Furthermore, due to the orthogonality of the DWT, the random errors in the wavelet domain remain independent and normally distributed with $\mathbb{E}(\varepsilon_i)=0$ and $ \mathrm{Var}(\varepsilon_i) = \sigma^2$. A wavelet shrinkage rule $\delta(\cdot)$ is applied to the empirical wavelet coefficients $\boldsymbol{d}$ to estimate $\boldsymbol{\theta}$, i.e $\boldsymbol{\hat{\theta}} = \delta(\boldsymbol{d})$, and the unknown function is estimated at the locations $x_i$ by the inverse discrete wavelet transform (IDWT) represented by $\boldsymbol{W'}$,
\begin{equation}
\boldsymbol{\hat{f}} = \boldsymbol{W'}\boldsymbol{\hat{\theta}}. \nonumber
\end{equation} 
For more details about wavelet shrinkage, see Vidakovic (1999). 

\begin{figure}[H]
\centering
\subfigure[Generated function $f$.]{
\includegraphics[scale=0.35]{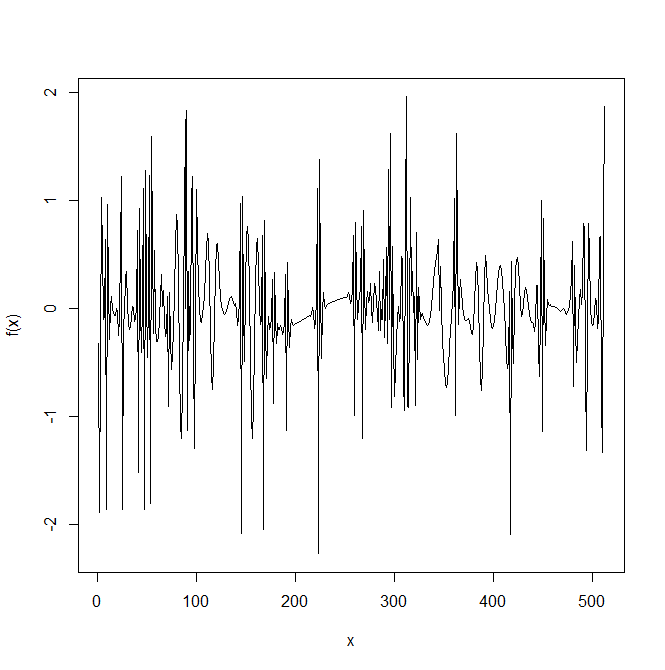}}
\subfigure[Generated dataset $\boldsymbol{y}$.]{
\includegraphics[scale=0.35]{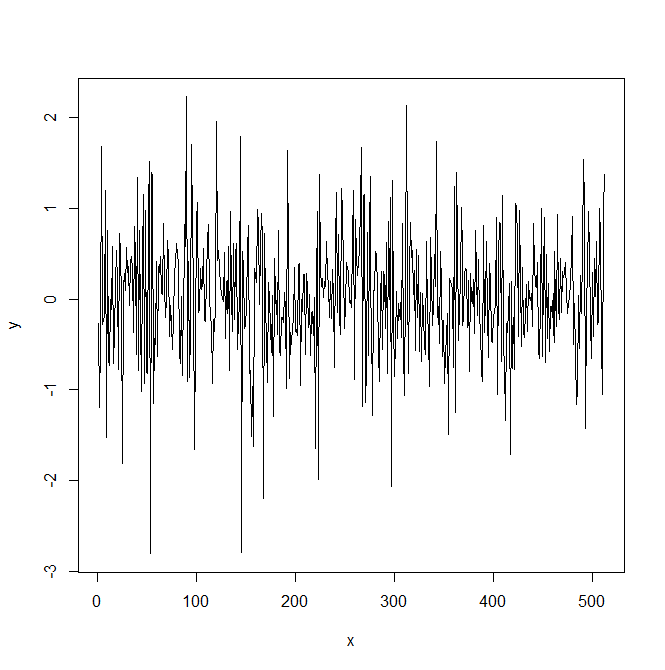}}
\subfigure[Estimated function $\hat{f}$.]{
\includegraphics[scale=0.35]{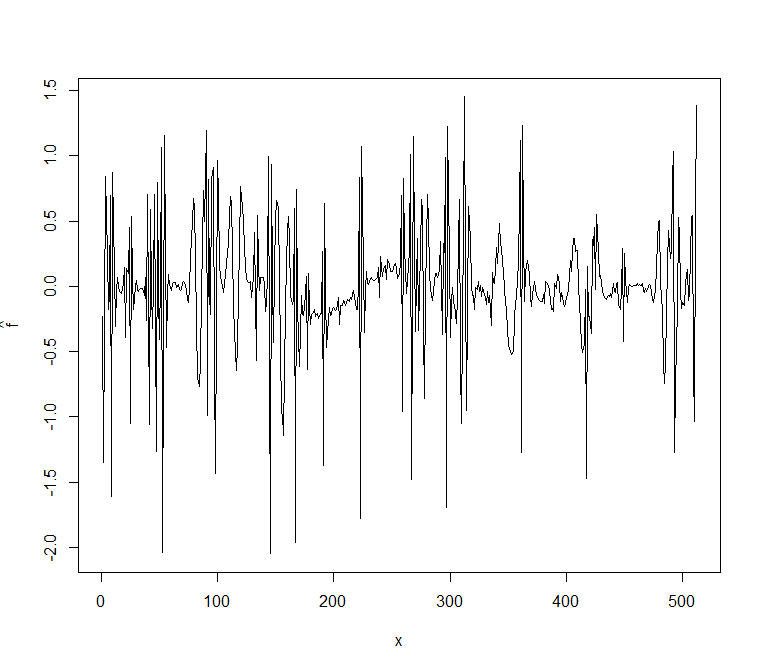}}
\caption{$n = 512$ points from the function $f$ (a), the generated dataset $\boldsymbol{y}$ (b) with asymmetrically distributed wavelet coefficients $\boldsymbol{\theta}$ and the estimated function $\hat{f}$ obtained by applying the Bayesian shrinkage rule under an asymmetric beta prior (c).} \label{fig:example}
\end{figure}

Our main concern here is to perform wavelet shrinkage when the coefficients are asymmetrically distributed. To illustrate this context, we generated a vector $\boldsymbol{f}$ with $n = 512$ points in such a way that its wavelet coefficients $\boldsymbol{\theta}$ were asymmetrically distributed according to a mixture of a point mass function at zero and an asymmetric beta distribution. We then generated the dataset $\boldsymbol{y}$ by adding random noise according to model \eqref{tmodel2}. The signal $\boldsymbol{f}$ and its noisy version $\boldsymbol{y}$ are available in Figures \ref{fig:example}(a) and (b) respectively. We propose a set of Bayesian shrinkage rules obtained from asymmetric priors with respect to $\boldsymbol{\theta}$ in order to estimate $f$ from $\boldsymbol{y}$. Figure \ref{fig:example}(c) shows the estimated function $\hat{f}$ by the application of one of the proposed Bayesian shrinkage rules, specifically obtained under the asymmetric beta prior to the wavelet coefficients.

\section{Bayesian shrinkage rules and their properties}

The Bayesian approach to perform wavelet shrinkage considers the sparsity of the vector of wavelet coefficients $\boldsymbol{\theta}$ and their support in the proposed prior distribution of the wavelet coefficients. Since the shrinkage rule is applied coefficient by coefficient, we consider a prior distribution $\pi(\cdot)$ to a single wavelet coefficient $\theta$ that is composed of a mixture of a point mass function at zero $\delta_0 (\theta)$ and an asymmetric distribution $g(\theta)$, i.e,
\begin{equation}\label{prior}
\pi(\theta) = \alpha \delta_0 (\theta) + (1-\alpha)g(\theta),
\end{equation}
where the weight $\alpha \in (0,1)$ is a hyperparameter that controls the shrinkage level of the rule. Higher values of $\alpha$ imply a greater level of shrinkage of the associated rule. Sousa (2022) proposed the asymmetric beta distribution as $g(\theta)$ in the prior \eqref{prior}. Here we propose a class $\Gamma$ of asymmetric distributions to be used as $g(\theta)$ in \eqref{prior} composed by 

\begin{figure}[H]
\centering
\subfigure[Beta densities.]{
\includegraphics[scale=0.4]{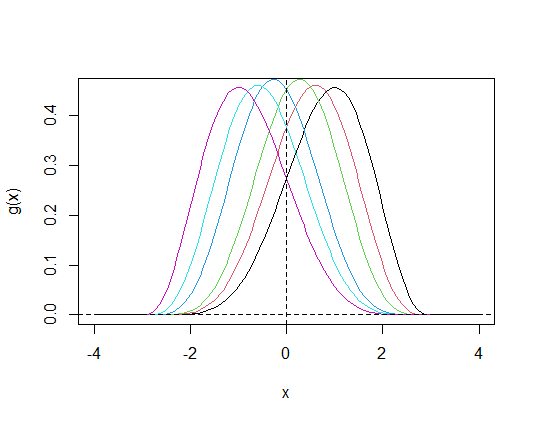}}
\subfigure[Kumaraswamy densities.]{
\includegraphics[scale=0.4]{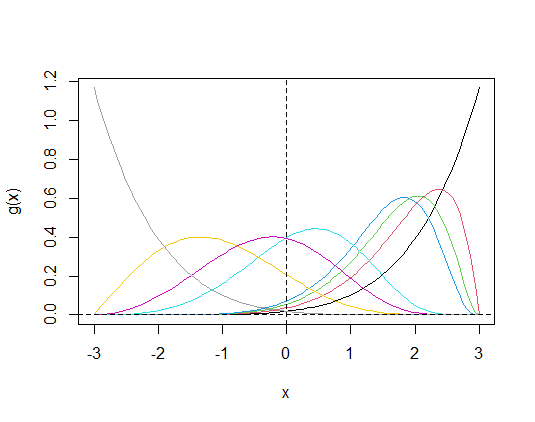}}
\subfigure[Triangular densities.]{
\includegraphics[scale=0.4]{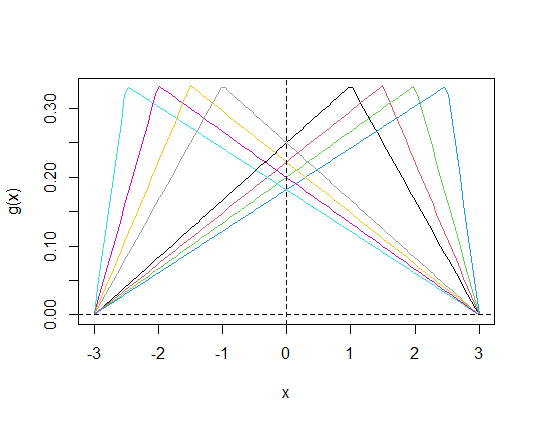}}
\subfigure[Skew normal densities.]{
\includegraphics[scale=0.4]{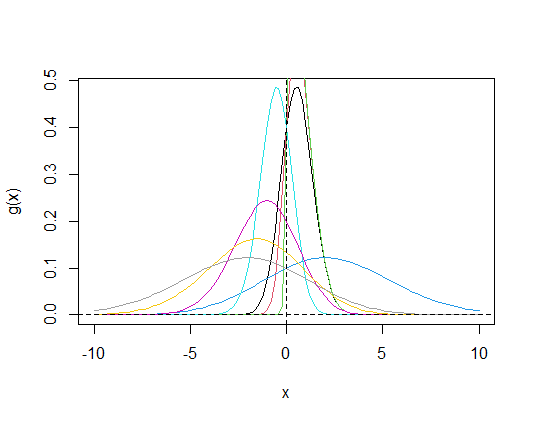}}
\caption{Distributions considered in the class $\Gamma$ of asymmetric distributions in the prior \eqref{prior} of wavelet coefficients for some values of their parameters: asymmetric beta (a), Kumaraswamy (b), asymmetric triangular (c) and skew normal (d).} \label{fig:dist}
\end{figure}

\begin{itemize}
\item the asymmetric beta distribution with probability density function $g_{B}(\cdot)$ given by
\begin{equation}\label{beta}
g_B (\theta)= g_B(\theta;a,b,m) = \frac{(\theta + m)^{a-1} (m - \theta)^{b-1}}{(2m)^{a+b-1}B(a,b)} \mathrm{I}_{(-m,m)}(\theta),
\end{equation}
where $a,b,m > 0$ and $a \neq b$, $B(a,b)$ is the beta function and $\mathrm{I}$ is the indicator function;

\item the Kumaraswamy distribution with probability density function $g_{K}(\cdot)$ given by
\begin{equation}\label{kum}
g_K(\theta) = g_K(\theta;a,b,m) = \frac{ab(\theta + m)^{a-1}[(2m)^a - (\theta + m)^a]^{b-1}}{(2m)^{ab}}\mathrm{I}_{(-m,m)}(\theta),
\end{equation}
where $a,b,m > 0$;

\item the asymmetric triangular distribution with probability density function $g_{T}(\cdot)$ given by
\begin{equation}\label{triang}
g_T (\theta) = g(\theta;a,m) = \frac{\theta + m}{m(m+a)} \mathrm{I}_{(-m,a]}(\theta) + \frac{m - \theta}{m(m-a)}\mathrm{I}_{(a,m)}(\theta),
\end{equation}
where $m > 0$, $a \in (-m,m)$ and $a \neq 0$;

\item and the skew normal distribution with probability density function $g_{S}(\cdot)$ given by
\begin{equation}\label{skewn}
g_S(\theta) = g_S(\theta;\tau,\gamma) = \frac{2}{\tau} \phi\left(\frac{\theta}{\tau} \right) \Phi\left(\frac{\gamma \theta}{\tau} \right) \mathrm{I}_{\mathbb{R}}(\theta),
\end{equation}
where $\tau > 0$, $\gamma \in \mathbb{R}$ and $\phi(\cdot)$ and $\Phi(\cdot)$ are the standard normal density and cumulative distribution functions respectively. We henceforth denote the proposed class of asymmetric distributions by $\Gamma = \{g_B, g_K, g_T, g_S \}$. Figure \ref{fig:dist} shows the densities \eqref{beta}, \eqref{kum}, \eqref{triang} and \eqref{skewn} for some of their hyperparameters.

\end{itemize}

The distributions that constitute the class $\Gamma$ were chosen to provide flexibility in the prior distribution to the wavelet coefficients in some aspects. The beta distribution as prior to $\theta$ was originally proposed by Sousa et al. (2020) but was restricted to the symmetric case ($a = b$), and later by Sousa (2022) considering the asymmetric case ($a \neq b$). Its well known that flexibility of shape makes the beta distribution  suitable for several coefficient distribution models. Furthermore, the associated shrinkage rule had good performance in denoising the empirical coefficients under small signal-to-noise ratios in the simulation studies. The Kumaraswamy distribution is a novelty in terms of priors to wavelet coefficients and was chosen due to its flexibility in shape and its closed relation with the beta distribution (for example if $X \sim \mathrm{Beta}(1,b)$, then $Y = X^{1/a} \sim \mathrm{Kum}(a,b)$, $a,b > 0$), although its density function does not depend on the beta function $B(a,b)$. This provides mathematical and computational advantages, see Jones (2009). This distribution was originally proposed by Kumaraswamy (1980) and since then it has been applied to several statistical modeling problems. 
The triangular distribution as prior to the wavelet coefficients was proposed by Sousa et al. (2020), but again only under the symmetric case. Its density function has a different shape in relation to the beta and Kumaraswamy densities, which expands the flexibility regarding the choice of the suitable prior. Finally, the skew normal distribution was chosen because its support is the real line, unlike the three first priors, which are bounded. Like the Kumaraswamy distribution, it is a novelty of the skew normal to be considered as prior to the wavelet coefficients.

\begin{figure}[H]
\centering
\subfigure[Beta rules.]{
\includegraphics[scale=0.45]{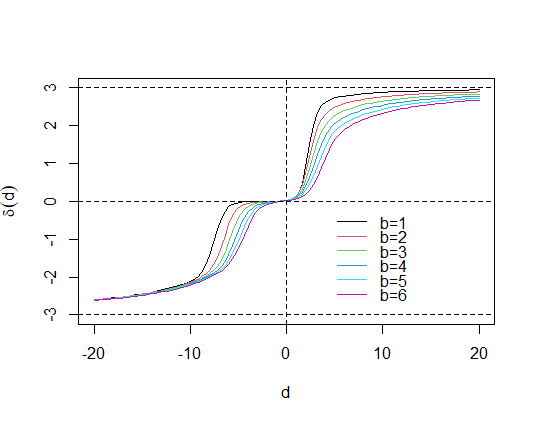}}
\subfigure[Kumaraswamy rules.]{
\includegraphics[scale=0.45]{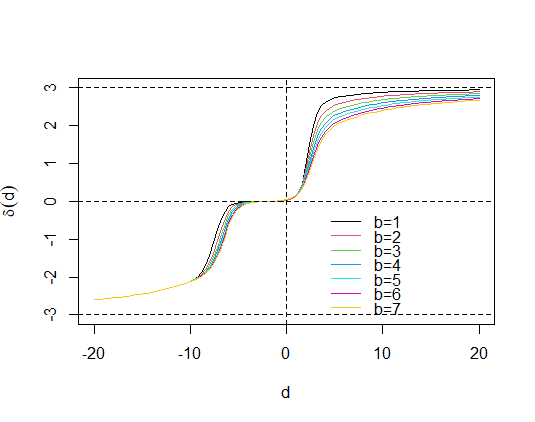}}
\subfigure[Triangular rules.]{
\includegraphics[scale=0.45]{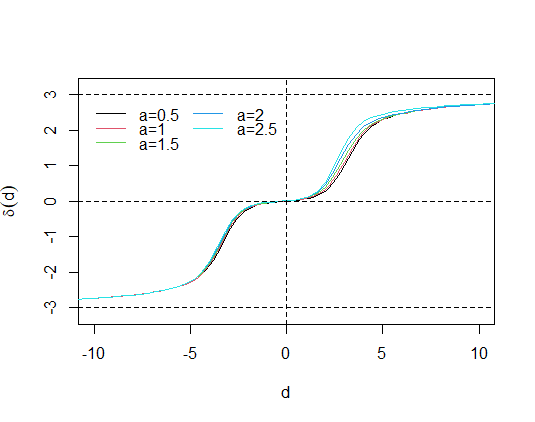}}
\subfigure[Skew normal rules.]{
\includegraphics[scale=0.45]{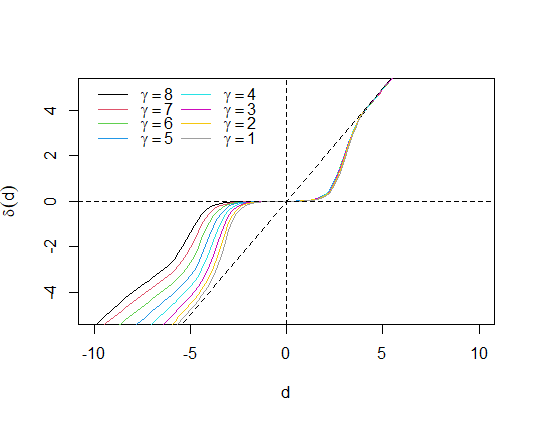}}
\caption{Bayesian shrinkage rules \eqref{rule} associated with the class $\Gamma$ of asymmetric distributions in the prior \eqref{prior} of wavelet coefficients for some values of their parameters, $\sigma = 1$, $\alpha = 0.9$ and $m=3$: asymmetric beta with $a=7$ (a), Kumaraswamy with $a=7$ (b), asymmetric triangular (c) and skew normal with $\tau=1$ (d).} \label{fig:rules}
\end{figure}

Under the squared error loss function $L(\delta,\theta) = (\delta - \theta)^2$, the Bayesian shrinkage rule is the posterior expected value of $\theta|d$. Sousa (2020) showed that under models \eqref{wmodel} and \eqref{prior}, the associated shrinkage rule is given by
\begin{equation}\label{rule}
\delta(d) = \mathbb{E}(\theta|d) = \frac{(1-\alpha)\int_\mathbb{R}(\sigma u + d)g(\sigma u +d)\phi(u)du}{\frac{\alpha}{\sigma}\phi(\frac{d}{\sigma})+(1-\alpha)\int_\mathbb{R}g(\sigma u +d)\phi(u)du},
\end{equation}
and it is obtained numerically. Figure \ref{fig:rules} shows the Bayesian shrinkage rules associated with the priors in the class $\Gamma$ for $\sigma = 1$, $\alpha = 0.9$ and $m = 3$. The hyperparameters of the distributions themselves were chosen to make them left asymmetric for simplicity. The interpretations of the right asymmetric case are similar. 

According to Figure \ref{fig:rules}, all the shrinkage rules reduce a sufficiently small empirical wavelet coefficient to zero or close to zero in an asymmetric way. Since the distributions of the class $\Gamma$ were chosen to be left asymmetric in these examples, the associated rules are more severe for empirical coefficients less than zero. On the other hand, for large empirical coefficients, the rules under priors with bounded support on $(-m,m)$ typically shrink them to values closed to $-m$ or $m$, i.e, the rules are bounded to the support of the prior. The interpretation is that since we have the prior information that $\theta \in (-m,m)$, then the exceeding value of the empirical coefficient occurs due the noise effect. 

The shrinkage rule under skew normal distribution is not bounded, since its support is the real line, although its behavior is different for large empirical coefficients greater or less than zero. In fact, the rules are more severe for coefficients less than zero and practically do not reduce the magnitudes of large coefficients greater than zero. For instance, if $d = 10$, this rule with hyperparameters $\gamma = \tau = 8$ (the black curve in Figure \ref{fig:rules}(d)) yields $\delta(10) = 9.8462$. However, if $d = -10$, then the same rule produces $\delta(-10) = -5.5202$ i.e, the rule shrinks the negative empirical coefficient more than the positive one with the same magnitude. 

Another important feature of the shrinkage rules is the impact of their hyperparameters on the shrinkage level. Small values of $b$ in the shrinkage rules under beta and Kumaraswamy priors and small values of $a$ and $\gamma$ in the rules under triangular and skew normal priors respectively imply a large shrinkage of the rules. This property should be taken into account in the elicitation of the hyperparameters of the selected rule. 

For the choices of $\alpha$ and $m$, we adopted the level-dependent proposals of Angelini and Vidakovic (2004),  
\begin{equation}\label{eq:alpha}
\alpha = \alpha(j) = 1 - \frac{1}{(j-J_{0}+1)^\beta},
\end{equation}
and
\begin{equation}\label{eq:m}
m = m(j) = \max_{k}\{|d_{jk}|\},
\end{equation}
where $J_ 0 < j \leq J-1$, $J_0$ is the primary resolution level, $J$ is the number of resolution levels, $J=\log_{2}(n)$ and $\beta > 0$. They also suggested that in the absence of additional information, $\beta = 2$ can be adopted. Finally, it is necessary to estimate $\sigma$ for complete (hyper)parameter elicitation. According to Donoho and Johnstone (1994), based on the fact that much of the noise information present in the data can be obtained on the finest resolution scale, they proposed
\begin{equation}\label{eq:sigma}
\hat{\sigma} = \frac{\mbox{median}\{|d_{J-1,k}|:k=0,...,2^{J-1}\}}{0.6745}.
\end{equation}

\begin{figure}[H]
\centering
\subfigure[Beta rules.]{
\includegraphics[scale=0.45]{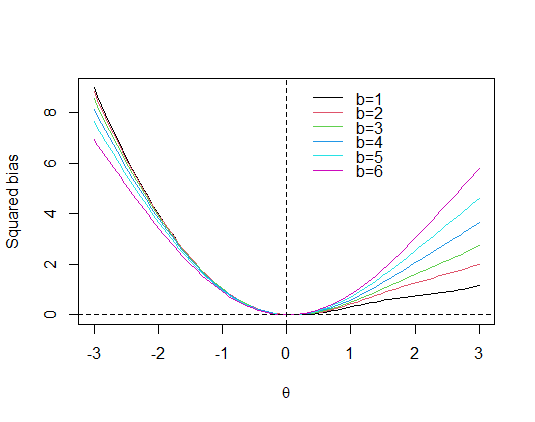}}
\subfigure[Kumaraswamy rules.]{
\includegraphics[scale=0.45]{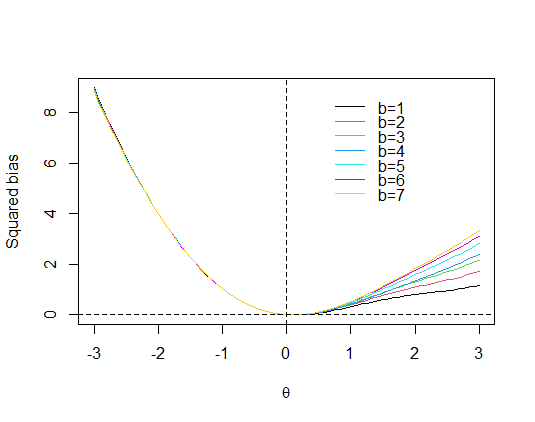}}
\subfigure[Triangular rules.]{
\includegraphics[scale=0.45]{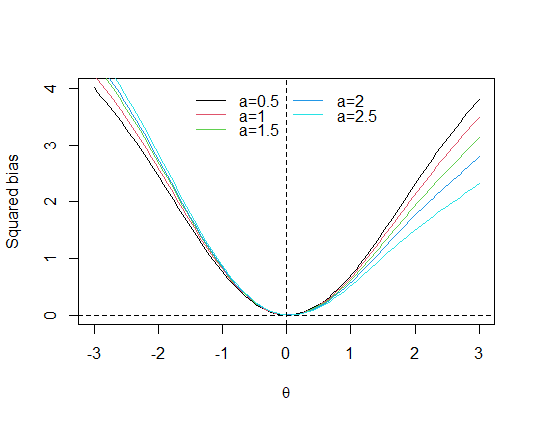}}
\subfigure[Skew normal rules.]{
\includegraphics[scale=0.45]{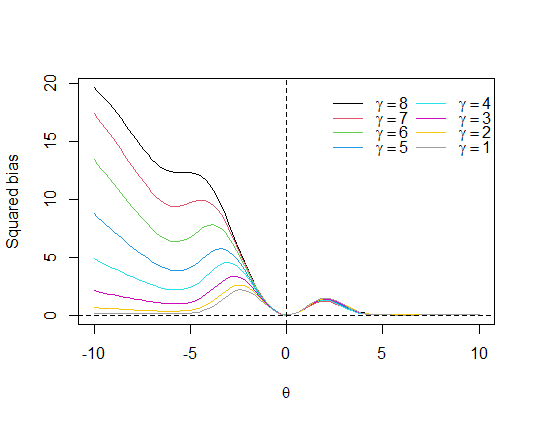}}
\caption{Squared bias of the Bayesian shrinkage rules \eqref{rule} associated to the class $\Gamma$ of asymmetric distributions in the prior \eqref{prior} of wavelet coefficients for some values of their parameters, $\sigma = 1$, $\alpha = 0.9$ and $m=3$: asymmetric beta with $a=7$ (a), Kumaraswamy with $a=7$ (b), asymmetric triangular (c) and skew normal with $\tau=1$ (d).} \label{fig:bias}
\end{figure}

The asymmetry of the prior distributions of $\Gamma$ is also reflected in the bias and  variances of their associated shrinkage rules. Figure \ref{fig:bias} shows the squared bias of the same shrinkage rules considered in Figure \ref{fig:rules}. In fact, the squared bias of the rules increases as $\theta$ increases in magnitude, but in  asymmetric fashion. The rules have higher squared bias when $\theta < 0$ than for $\theta > 0$. For example, the squared bias of the shrinkage rule under an asymmetric beta prior with hyperparameters $a=7$ and $b=1$ (the black curve in Figure \ref{fig:bias}(a)) when $\theta = -2$ is equal to 4.0071 while for $\theta = 2$ it is equal to 0.8828. This characteristic of the bias is reasonable since the considered rule shrinks the empirical coefficients $d < 0$ more, so its expected value is small, which implies high bias when the true wavelet coefficient is significant.

\begin{figure}[H]
\centering
\subfigure[Beta rules.]{
\includegraphics[scale=0.45]{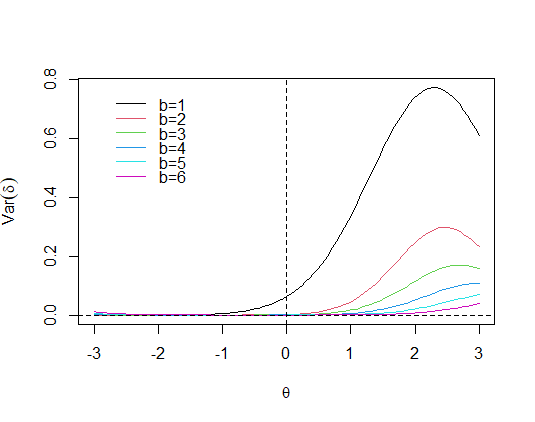}}
\subfigure[Kumaraswamy rules.]{
\includegraphics[scale=0.45]{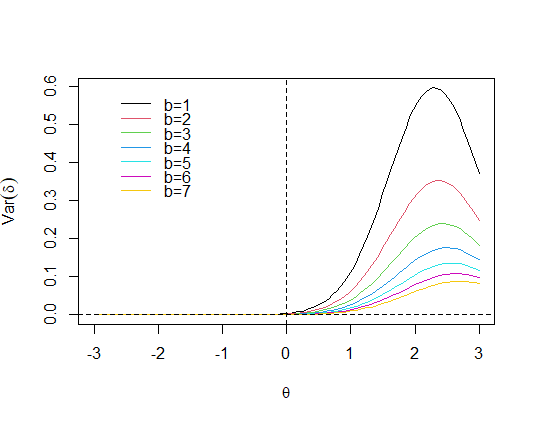}}
\subfigure[Triangular rules.]{
\includegraphics[scale=0.45]{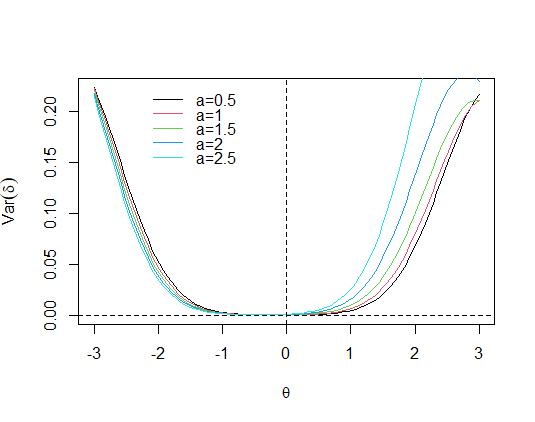}}
\subfigure[Skew normal rules.]{
\includegraphics[scale=0.45]{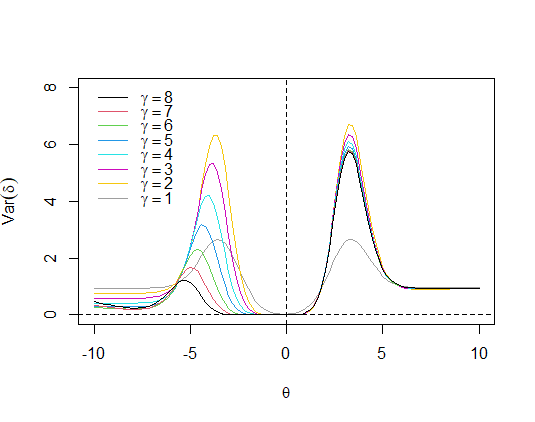}}
\caption{Variances of the bayesian shrinkage rules \eqref{rule} associated to the class $\Gamma$ of asymmetric distributions in the prior \eqref{prior} of wavelet coefficients for some values of their parameters, $\sigma = 1$, $\alpha = 0.9$ and $m=3$: asymmetric beta with $a=7$ (a), Kumaraswamy with $a=7$ (b), asymmetric triangular (c) and skew normal with $\tau=1$ (d).} \label{fig:var}
\end{figure}

The variances of the shrinkage rules are provided in Figure \ref{fig:var} and have opposite behaviour in relation to the squared bias, as expected. Actually the variances are higher for $\theta > 0$ than for $\theta < 0$. For instance, the same shrinkage rule under asymmetric beta prior considered in the example of the bias given previously, its variance is equals to 0.0002 when $\theta = -2$ and it is equals to 0.7390 for $\theta = 2$. Further, the hyperparameters also impact on the variance of the rules, which is expected since they determine the shrinkage level of the rules. As an illustration, the shrinkage rule under asymmetric beta prior with $b = 3$ has variance equals to 0.3366 when $\theta = 2$.

Finally, Table \ref{tab:brisk} provides the Bayesian risks of the shrinkage rules under the priors in $\Gamma$ in relation to each prior for the (hyper)parameters $\sigma = 1$, $\alpha = 0.9$, $m = 3$ and $(a = 7;b=2)$ for the asymmetric beta, $(a= 7, b = 1)$ for the Kumaraswamy, $a = 1$ for the triangular and $(\tau = 8;\gamma = 3)$ for the skew normal. In fact, under the asymmetric beta, Kumaraswamy and the triangular priors, the risks of the four shrinkage rules are closed among themselves, which denotes evidence of robustness in terms of the choice of the prior in $\Gamma$ in relation to these priors. As expected, under the asymmetric beta and the triangular priors, the shrinkage rule under the Kumaraswamy prior has the smallest risk (we do not consider the shrinkage rule under asymmetric beta prior) and under the Kumaraswamy prior, the beta shrinkage rule has the smallest risk. 

On the other hand, under the skew normal prior, the risks are rising. For instance, the shrinkage rule under skew normal prior has risk equal to 0.1750 while the main risk (the rule under asymmetric beta prior) is equal to 3.4608. Hence, these results suggest that the shrinkage rule under the skew normal distribution might be suitable for the other priors in $\Gamma$ while the inverse is not true, i.e, for the skew normal prior, the rules under beta, Kumaraswamy and triangular setups are not as suitable in terms of Bayesian risks, which is reasonable due the difference in the supports of the skew normal and the others. More about robustness from Bayesian point of view can be seen in Berger (1984, 1990).

\begin{table}[H]
\centering
\caption{Bayesian risks of the shrinkage rules \eqref{rule} determined by the priors in $\Gamma$ with respect to each prior. The chosen hyperparameters are $\alpha = 0.9$, $m = 3$ and $(a = 7;b=2)$ for the asymmetric beta, $(a=7,b=1)$ for the Kumaraswamy, $a = 1$ for the triangular and $(\tau = 8;\gamma = 3)$ for the skew normal. Furthermore, it was considered $\sigma = 1$.}\label{tab:brisk}
\begin{tabular}{|c|c|c||c|c|c|}
\hline
\textbf{Prior} & \textbf{Rule} & \textbf{Bayesian risk} & \textbf{Prior} & \textbf{Rule} & \textbf{Bayesian risk} \\ \hline
Beta & Beta & 0.1893             & Kumaraswamy & Beta & 0.2282 \\
     & Kumaraswamy & 0.2043    &        & Kumaraswamy & 0.2135 \\
     & Triangular & 0.2078     &        & Triangular & 0.3172  \\
     & Skew normal & 0.3235    &        & Skew normal & 0.2966 \\
 \hline \hline
Triangular & Beta & 0.2133     &  Skew normal & Beta & 3.4608 \\
           & Kumaraswamy & 0.1527   &    & Kumaraswamy & 4.0127 \\
           & Triangular & 0.1209    &    & Triangular & 4.0078 \\
           & Skew normal & 0.1901   &    & Skew normal & 0.1750 \\ \hline
\end{tabular}
\end{table}

\section{Simulation studies}

The performances of the shrinkage rules \eqref{rule} were obtained in two Monte Carlo simulation studies. 

We chose the shrinkage rules under the asymmetric beta prior with $a = 5$ and $b=1$ (ASYBETA), Kumaraswamy prior with $a = 7$ and $b = 2$ (KUM), triangular prior with $a = 8$ (TRI) and skew normal prior with $\tau = 8$ and $\gamma = 4$ (SN). Moreover, we compared their performances against the shrinkage rule under the symmetric beta prior with parameters $a = b = 5$ (SYMBETA) proposed by Sousa et al. (2020), the large posterior mode shrinkage rule (LPM) described by Cutillo et al. (2008) and the standard soft thresholding rule proposed by Donoho and Johnstone (1994) with two policies to choose the threshold value, the cross-validation policy (CV) by Nason (1996) and the Stein unbiased risk estimator (SURE) proposed by Donoho and Johnstone (1995). Since all these rules were developed under the symmetric assumption of the distribution of the wavelet coefficients, we expected the proposed asymmetric rules outperform these ones.

The performance measures considered were the averaged mean squared error (AMSE), 
\begin{equation}
\mathrm{AMSE} = \frac{1}{Rn}\sum_{r = 1}^{R}\sum_{i = 1}^{n}[\hat{f}_r(x_i) - f(x_i)]^2, \nonumber
\end{equation}
where $\hat{f}_r(x_i)$ is the estimate of $f(x_i)$ in the replication $r$, and the averaged median absolute error (AMAE)
\begin{equation}
\mathrm{AMAE} = \frac{1}{R}\sum_{r = 1}^{R} \mathrm{Med}\{|\hat{f}_r(x_i) - f(x_i)|, i = 1,\cdots,n \}, \nonumber
\end{equation}
where $\mathrm{Med}$ denotes the median. We considered $R = 200$ replications in both simulation studies.

\subsection{Simulation study 1}

In this first study, we generated asymmetric wavelet coefficients according to the prior distribution \eqref{prior} and added random noises according to model \eqref{wmodel} in the wavelet domain with two sample sizes, $n = 512$ and $2048$. We chose $g(\theta)$ to be the asymmetric beta density \eqref{beta} with parameters $m = 10$, $a = 7$ and $b = 1$ (scenario 1 - weak asymmetry) and $a = 20$ and $b = 1$ (scenario 2 - strong asymmetry). The variance $\sigma^2$ was chosen according to two signal-to-noise ratios ($\mathrm{SNR}$), $\mathrm{SNR} = 3$ and $9$.

Table \ref{tab:sim1} presents the results of scenario 1 of weak asymmetry of the wavelet coefficients. In general the performance of the rules was strongly influenced by the signal-to-noise ratio, i.e, the performances were better for $\mathrm{SNR} = 9$ than for $\mathrm{SNR} = 3$, as expected. On the other hand, the sample size did not greatly impact their performances. No improvements in the AMSE and AMAE values were observed for $n = 2048$ against $n = 512$. For instance, the triangular shrinkage rule had AMSE equal to 0.0523 for $n=512$ and $\mathrm{SNR} = 3$, which declined to 0.0074 for $\mathrm{SNR} = 9$ and the same sample size, an approximate 7-fold reduction. On the other hand its AMSE was 0.0798 for $n = 2048$ and $\mathrm{SNR} = 3$, which is almost the same as for $n = 512$ and the same SNR.  

Comparison of the rules themselves revealed that the proposed asymmetric rules outperformed the standard ones in all the combinations of signal-to-noise ratio and sample size, which also was expected since the standard rules were developed under symmetric noises. SURE and the symmetric beta were the best among the standard techniques but with AMSE and AMAE values larger than the asymmetric rules values. For instance, the AMSE of SURE was equal to 0.1511 and the AMSE of the symmetric beta rule was 0.0854 against AMSE equal to 0.0461 for the Kumaraswamy rule when $\mathrm{SNR} = 3$ and $n = 512$.  In general, the asymmetric beta shrinkage rule had the best performance in scenario 1, but with no significant difference against the other asymmetric rules. It should also be noted that the standard thresholding rule under the CV policy did not perform well in general. For example, for $n = 2048$ and $\mathrm{SNR} = 3$, the AMSE was 1.3232 against 0.0611 for the asymmetric beta rule, i.e, about 21 times bigger than this rule's values.

\begin{figure}[H]
\centering
\subfigure[Scenario 1 - MSE.]{
\includegraphics[scale=0.3]{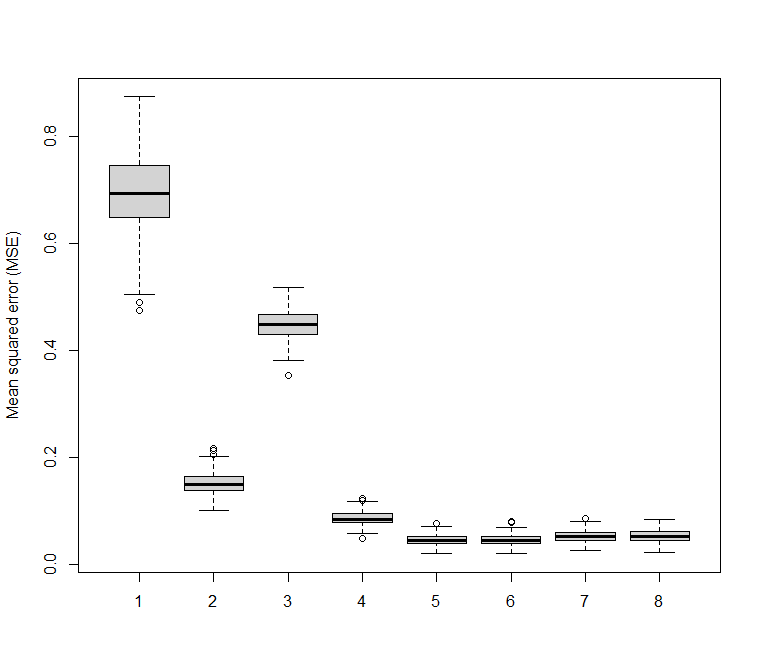}}
\subfigure[Scenario 1 - MAE.]{
\includegraphics[scale=0.3]{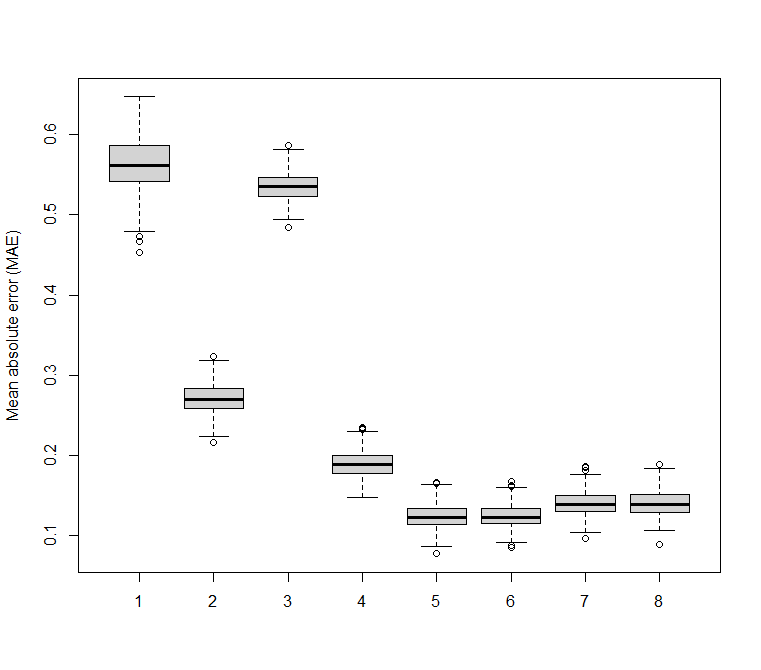}}
\subfigure[Scenario 2 - MAE.]{
\includegraphics[scale=0.3]{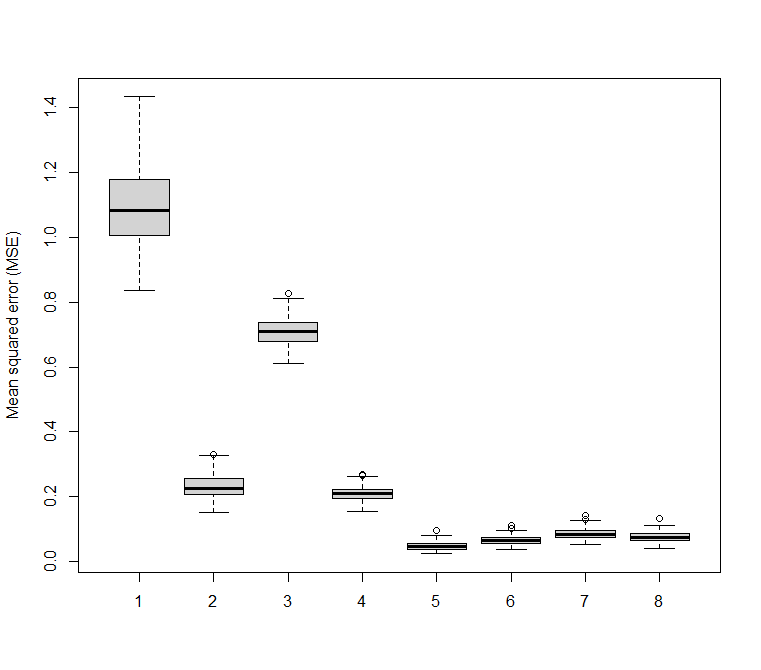}}
\subfigure[Scenario 2 - MAE.]{
\includegraphics[scale=0.3]{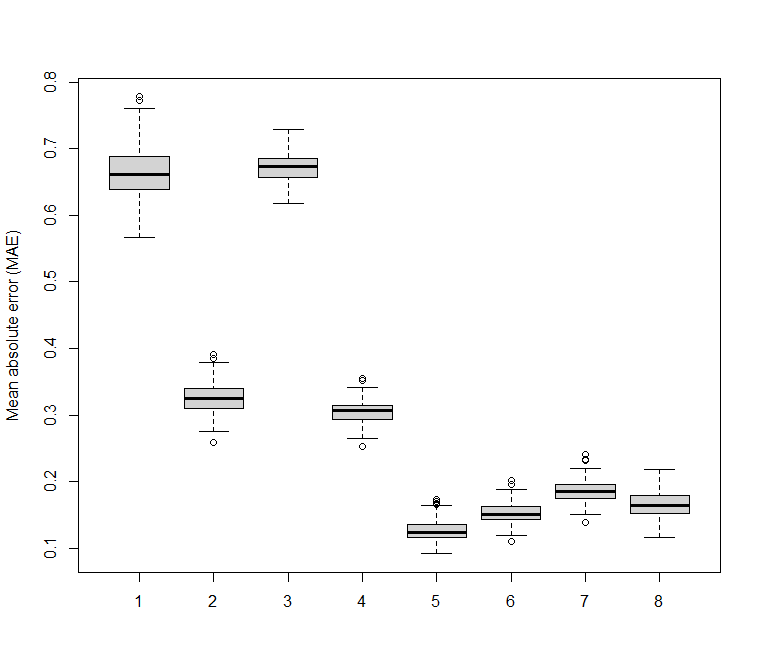}}

\caption{Boxplots of the MSE and MAE values of the shrinkage and thresholding rules considered in the simulation study 1 in the scenario 1 - MSEs (a), scenario 1 - MAE (b), scenario 2 - MSE (c) and scenario 2 - MAE (D). In all the cases with $n = 512$ and 
$\mathrm{SNR} = 3$. The associated rules are: 1-CV, 2-SURE, 3-LPM, 4-BETASYM, 5-BETAASYM, 6-KUM, 7-TRI and 8-SN.} \label{fig:bp}
\end{figure}

Although the general behavior of the performance measures was the same, the AMAE values were bigger than the AMSE values according to nearly all the methods. For $n=512$ and $\mathrm{SNR} = 9$, the asymmetric beta had AMSE equal t 0.0069 but AMAE equal to 0.0505. Finally, small dispersions of the MSE and MAE values of the proposed rules can be seen in Figures \ref{fig:bp}(a) and \ref{fig:bp}(b), which show the boxplots of the MSE and MAE values under $\mathrm{SNR} = 3$ and $n = 512$ respectively for scenario 1, confirming the generally good behavior of the asymmetric rules in this scenario.

\begin{table}[H]
\centering
\label{my-label}
\begin{tabular}{|c|c|c|c|}
\hline
$\boldsymbol{n}$ & \textbf{Method} & $\boldsymbol{\mathrm{SNR = 3}}$ & $\boldsymbol{\mathrm{SNR = 9}}$ \\ \hline \hline
& & \textbf{AMSE (SD) AMAE (SD)} & \textbf{AMSE (SD) AMAE (SD)} \\ \hline \hline
512	&	CV	& 0.6937 (0.0759) 	0.5627 (0.0332)	& 0.1247 (0.0149) 0.2406 (0.0149)	\\
	&	SURE	&	0.1511 (0.0218) 0.2707 (0.0192)	& 0.0242	 (0.0037) 0.1083 (0.0079)	\\
	&	LPM	&	0.4485	(0.0275) 0.5349 (0.0177)	& 0.0650	 (0.0043) 0.2035 (0.0072)	\\
	&	BETASYM	& 0.0854	 (0.0132) 0.1892 (0.0162)	& 0.0109	 (0.0043) 0.0652 (0.0071)\\
	&	BETAASYM	&	\textbf{0.0458	(0.0104)} \textbf{0.1243 (0.0153)}	& \textbf{0.0069	 (0.0015)} \textbf{0.0505 (0.0062)}	\\
	&	KUM	&	0.0461	(0.0103) 0.1243 (0.0148)	& 0.0070	 (0.0015) 0.0506	 (0.0061)	\\
	&	TRI	& 0.0523	 (0.0109) 0.1397	 (0.0155)	&0.0074 (0.0016) 0.0533(0.0063)		\\
	&	SN	&0.0528	(0.0113)	0.1397 (0.0162)	&0.0075	(0.0016)	 0.0536 (0.0064)	\\ \hline \hline
									
2048	&	CV	&	1.3232 (0.0719)	0.7718 (0.0217)	& 0.1482	 (0.0093) 0.2559 (0.0083)	\\
	&	SURE	& 0.2158	 (0.0170) 0.3148	 (0.0112)& 0.0237 (0.0019) 0.1035 (0.0038)			\\
	&	LPM	& 0.6040	 (0.0184) 0.6199	 (0.0098)	&0.0668	(0.0020) 0.2061	(0.0034)	\\
	&	BETASYM	& 0.1522	 (0.0094) 0.2564	 (0.0086)	& 0.0117	 (0.0010)	0.0662 (0.0031)	\\
	&	BETAASYM	& \textbf{0.0611	 (0.0072)} \textbf{0.1469 (0.0084)} &	\textbf{0.0066 (0.0007) 0.0480 (0.0028)}\\
	&	KUM	& 0.0663	 (0.0073) 0.1552	 (0.0081)	&	0.0070 (0.0007)	0.0496 (0.0028)	\\
	&	TRI	&0.0798	(0.0077)	 0.1780 (0.0086) &	0.0076 (0.0007) 0.0496 (0.0029)	\\
	&	SN	&0.0767	(0.0076) 0.1717	(0.0085)	&	0.0074 (0.0007) 0.0515 (0.0029)		\\ \hline

\end{tabular}
\caption{AMSE, AMAE and their standard deviations (SD) of the shrinkage and thresholding rules considered in the simulation study 1 for the scenario 1 (weak asymmetry of the wavelet coefficients).}\label{tab:sim1}
\end{table}

The results of scenario 2 (strong asymmetry) are reported in Table \ref{tab:sim2}. The general performance of the methods under this scenario was slightly worse than under scenario 1. For example, the AMSE of the triangular shrinkage rule was equal to 0.0798 for $\mathrm{SNR = 3}$ and $n = 2048$ in scenario 1 but equal to 0.0870 in scenario 2 for the same context of signal-to-noise ratio and sample size. As occurred in scenario 1, the proposed asymmetric rules performed well and were very similar among themselves, i.e, although the asymmetric beta rule had the best performance, the other asymmetric rules had AMSE and AMAE values close to the asymmetric beta values.
Moreover, the Kumaraswamy shrinkage rule performed better than the others for $n=2048$ and $\mathrm{SNR = 9}$ in terms of AMAE. Actually the triangular and skew normal rules outperformed the asymmetric beta rule under these sample size and SNR values when considering the AMAE. 

 Figure \ref{fig:bp}(c) and Figure \ref{fig:bp}(d) contain  boxplots of the MSE and AMAE values for $\mathrm{SNR} = 3$ and $n = 512$ of scenario 2 respectively. As in scenario 1, the proposed rules had small dispersion in MSE and AMAE values with relation to the standard ones. Similar results were obtained for the other sample size and SNR values. 

Thus the general performance of the asymmetric rules was better than that of the standard rules in both scenarios of asymmetry. Furthermore, although the asymmetric beta were the best in almost all the combinations of sample size and SNR, which is reasonable since the wavelet coefficients were generated under the asymmetric beta distribution, the other asymmetric rules had similar performance to that one in terms of AMSE and AMAE. This indicates that the choice for a specific prior of the proposed class should be done mainly according to the believed shape of the prior distribution of the wavelet coefficients, since the associated rules had similar performance in terms of precision.

\begin{table}[H]
\centering
\label{my-label}
\begin{tabular}{|c|c|c|c|}
\hline
$\boldsymbol{n}$ & \textbf{Method} & $\boldsymbol{\mathrm{SNR = 3}}$ & $\boldsymbol{\mathrm{SNR = 9}}$ \\ \hline \hline
& & \textbf{AMSE (SD) AMAE (SD)} & \textbf{AMSE (SD) AMAE (SD)} \\ \hline \hline

512	&	CV	&1.0947	(0.1247) 	0.6616 (0.0387)	&0.1654	(0.0097) 0.2679 (0.0170)	\\
	&	SURE	&0.2307 (0.0339) 0.3247	(0.0218)		&0.0321	(0.0045) 0.1235 (0.0086)		\\
	&	LPM	&0.7093	(0.0427) 	0.6719 (0.0212)	&0.0882 (0.0052)	0.2368 (0.0061)	\\
	&	BETASYM	&0.2095	(0.0214) 0.3053 (0.0163)		&	0.0201 (0.0024) 0.0894	 (0.0061)	\\
	&	BETAASYM	&\textbf{0.0479	(0.0119) 0.1261 (0.0158)}		&\textbf{0.0072	(0.0014) 0.0491	(0.0053)}	\\
	&	KUM	&0.0647	(0.0128)	 0.1525 (0.0153)	&0.0084	(0.0015)	0.0540 (0.0052)	\\
	&	TRI	&0.0870	(0.0154)	 0.1860 (0.0161)	&0.0097	(0.0016)	0.0598 (0.0055)	\\
	&	SN	&0.0754	(0.0154)	 0.1651 (0.0194)		&0.0097	(0.0016) 0.0574 (0.0057)	\\ \hline
																	
2048	&	CV	&1.4973	(0.0924) 0.7984 (0.0253) 		&	0.2056 (0.0112) 0.2956	 (0.0084)\\
	&	SURE	&0.2511	(0.0192)	 0.3341 (0.0125)	&0.0324 (0.0025)	0.2956 (0.0041)	\\
	&	LPM	&0.7516	(0.0252) 0.6915 (0.0122)		&0.0908 (0.0026) 0.2956 (0.0041)		\\
	&	BETASYM	&0.2388	(0.0115) 	0.3282 (0.0091)	&0.0212	(0.0014) 0.0924 (0.0034)	\\
	&	BETAASYM	&\textbf{0.0517	(0.0067)	 0.1339 (0.0083)}	&\textbf{0.0076 (0.0007)} 0.0924 (0.0028)	\\
	&	KUM	&0.0715	(0.0071)	 0.1632 (0.0083)&	0.0090 (0.0008) \textbf{0.0557 (0.0028)}		\\
	&	TRI	&0.0963	(0.0084)	 0.1982 (0.0088)	&0.0105 (0.0009)	0.0616 (0.0028)		\\
	&	SN	&0.0842	(0.0089)	 0.1775 (0.0102)	&0.0099 (0.0009) 0.0616 (0.0031)			\\ \hline

\end{tabular}
\caption{AMSEs, AMAEs and their standard deviations (SD) of the shrinkage and thresholding rules considered in the simulation study 1 for scenario 2 (strong asymmetry of the wavelet coefficients).}\label{tab:sim2}
\end{table}

\subsection{Simulation study 2}

\begin{figure}[H]
\centering
\includegraphics[scale=0.80]{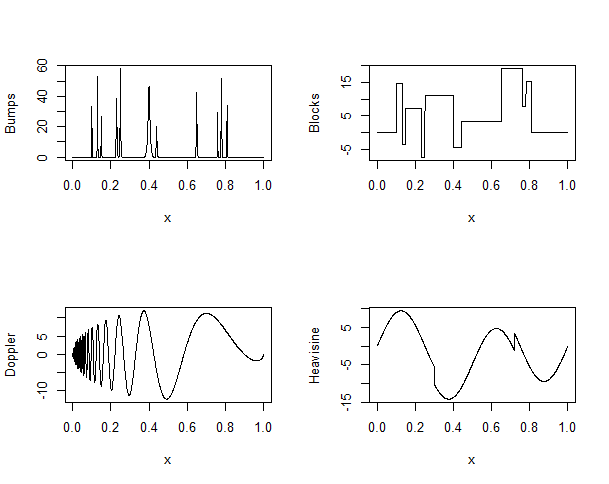}
\caption{Donoho and Johnstone test functions used as underlying functions in simulation study 2.}\label{fig:dj}
\end{figure}

The second simulation study involves the four Donoho and Johnstone test functions called Bumps, Blocks, Doppler and Heavisine as underlying functions in the time model \eqref{tmodel} for $x \in [0,1]$. These functions are broadly applied in the literature to evaluate wavelet-based methods in nonparametric regression models, since they have several interesting features to estimate. For instance, Bumps is composed of spikes, Blocks has type 2 discontinuities (jumps), Doppler has oscillations with different frequencies and Heavisine has a cusp. The plots of the four functions are provided in Figure \ref{fig:dj}. 

\begin{figure}[H]
\centering
\subfigure[$\mathrm{SNR} = 3$]{
\includegraphics[scale=0.35]{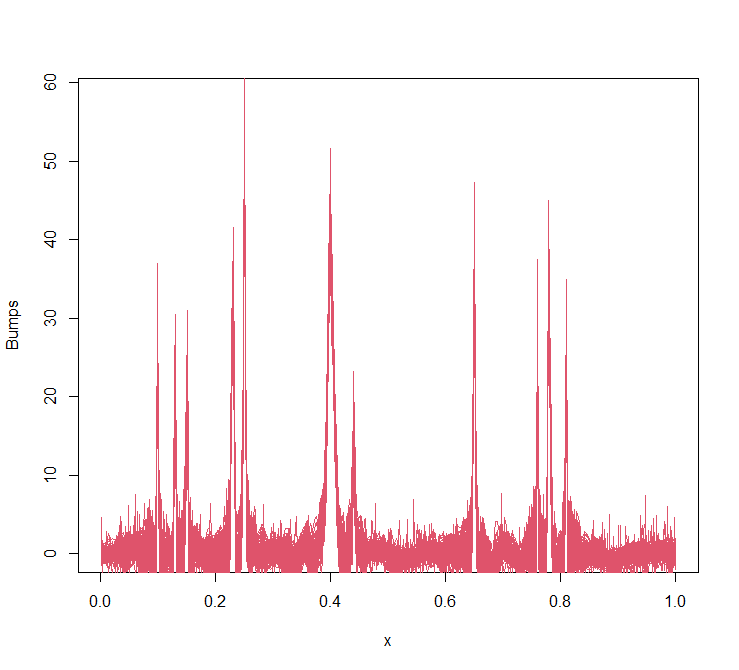}}
\subfigure[$\mathrm{SNR} = 9$]{
\includegraphics[scale=0.35]{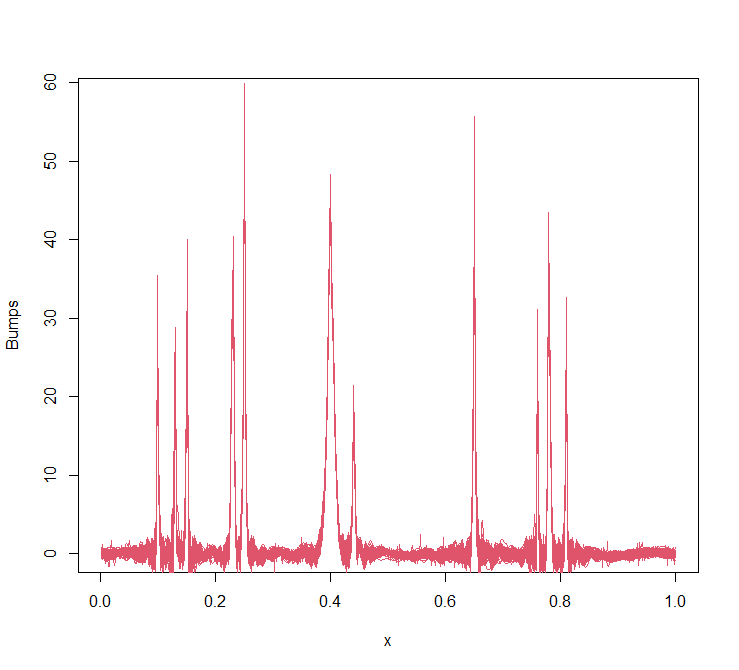}}

\caption{$M = 200$ Bumps estimates by the asymmetric triangular shrinkage rule for $n = 512$, $\mathrm{SNR}=3$ (a) and $\mathrm{SNR}=9$ (b) in the simulation study 2.} \label{fig:bumps}
\end{figure}

For each test function, points were generated according to \eqref{tmodel} for $n = 512$ and $2048$ and for $\sigma^2$ according to $\mathrm{SNR} = 3$ and $9$, as in the simulation study 1. The difference of this study in relation to the first one is that the wavelet coefficients of the test functions are not necessarily asymmetrically distributed. Table \ref{tab:sim3} shows the AMSEs and AMAEs of the rules for Bumps and Blocks underlying functions and Table \ref{tab:sim4} presents the results of the Doppler and Heavisine test functions. In general the rules had similar performance for the four test functions. The asymmetric triangular shrinkage rule was the best one in terms of both measures for Bumps, which indicates the asymmetry of its coefficients. The triangular rule was also the best for Doppler with $\mathrm{SNR} = 3$. Although the triangular rule outperformed the other methods in those scenarios, the asymmetric beta and Kumaraswamy rules also worked well in. For instance, the AMSE of the triangular rule was 1.2124 for Bumps with $n = 2048$ and $\mathrm{SNR} = 3$ versus 1.2734 for the asymmetric beta and Kumaraswamy rules. The same behavior occurred for the AMAEs of these rules. Figure \ref{fig:bumps} presents the $M = 200$ estimates obtained by the asymmetric triangular rule for Bumps with $n = 512$ and $\mathrm{SNR} =3$ (a) and  $\mathrm{SNR} =9$ (b) respectively. The  estimates were better for $\mathrm{SNR} =9$ than  $\mathrm{SNR} =3$ as expected, but even under large presence of noise in the data such as for $\mathrm{SNR} =3$, the estimates performed well in identifying the spikes, which are the main characteristics of Bumps. 

On the other hand, the symmetric rules outperformed the asymmetric ones for the Blocks and Heavisine functions, which indicates a symmetric distribution of their wavelet coefficients. For Blocks, the symmetric beta rule and SURE method were the best in both measures, while the CV method was the best in all the sample size and SNR scenarios for the Heavisine function. However even for these underlying functions, the asymmetric triangular, beta and Kumaraswamy rules had  good performance. For example, the AMSE of CV was 0.5153 for Heavisine with $n = 512$ and $\mathrm{SNR} = 3$ versus 0.5376, 0.5376 and 0.5385 for the asymmetric beta, Kumaraswamy and triangular rules respectively. Similar comparisons can also be made for their AMAEs.

The skew normal shrinkage rule and the LPM did not perform well in all the scenarios. 

\begin{table}[H]
\centering
\label{my-label}
\begin{tabular}{|c|c|c|c|c|}
\hline
Signal & $\boldsymbol{n}$ & \textbf{Method} & $\boldsymbol{\mathrm{SNR = 3}}$ & $\boldsymbol{\mathrm{SNR = 9}}$ \\ \hline \hline
&& & \textbf{AMSE (SD) AMAE (SD)} & \textbf{AMSE (SD) AMAE (SD)} \\ \hline \hline

Bumps & $512$ &CV&11.5102	(1.4661)	1.7886	(0.1003)	&	4.5341	(0.3401)	1.2150	(0.0424) \\
&&SURE &3.7231	(0.5152)	1.2306	(0.0613)	&	0.5746	(0.0804)	0.5088	(0.0293) \\
&&LPM&5.4702 (0.3248)	1.8659	(0.0594)	&	0.6054	(0.0338)	0.6201	(0.0186) \\
&&BETASYM&\textbf{2.8159	(0.2708)	1.1743	(0.0665)}	&	0.4439	(0.1341)	0.4786	(0.0447) \\
&&BETAASYM&3.0200	(0.2805)	1.2209	(0.0668)	&	0.5160	(0.3976)	0.4942	(0.0378)\\
&&KUM&3.0200(	0.2805)	1.2209	(0.0668)	&	0.5160	(0.3976)	0.4942	(0.0378)\\
&&TRI&2.8629	(0.2736)	1.1901	(0.0681)	&	\textbf{0.4282	(0.0383)	0.4730	(0.0230)}\\
&&SN&7.9392	(0.4077)	1.7397	(0.0608)	&	0.6860	(0.0427)	0.5558	(0.0220)\\ \hline
								
& $2048$ &CV&1.6139	(0.1055)	0.8201	(0.0305)	&	0.3833	(0.0254)	0.3799	(0.0148) \\
&&SURE&1.6694	(0.1257)	0.8217	(0.0313)	&	0.2530	(0.0196)	0.3300	(0.0116) \\
&&LPM&5.4483(	0.1725)	1.8627	(0.0309)	&	0.6071	(0.0186)	0.6221	(0.0102) \\
&&BETASYM&1.2125	(0.0959)	0.7461	(0.0357)	&	1.5863	(1.1849)	0.7213	(0.3566) \\
&&BETAASYM&1.2734	(0.1031)	0.7626	(0.0382)	&	0.2431	(0.3241) \textbf{	0.3139	(0.0261)} \\
&&KUM&1.2734(	0.1031)	0.7626	(0.0382)	&	0.2431	(0.3241) \textbf{	0.3139	(0.0261)} \\
&&TRI&\textbf{1.2124	(0.0970)	0.7448	(0.0364)}	&	\textbf{0.2369	(0.2690)}	0.3190	(0.0261) \\
&&SN& 9.3185 (1.2009) 2.0761 (0.2644)	&	0.3547 (0.0172) 0.3841 (0.0123) \\ \hline \hline

Blocks&512	&	CV	&	2.5672	(0.2628)	\textbf{1.1504	(0.0683)}	&	0.6721	(0.0596)	0.5615	(0.0307) \\
&	&	SURE	&	2.8156	(0.3915)	1.1890	(0.0834)	&	\textbf{0.4512	(0.0673)	0.4725	(0.0312)} \\
&	&	LPM	&	5.4532	(0.3500)	1.8601	(0.0614)	&	0.6060	(0.0364)	0.6215	(0.0211) \\
&	&	BETASYM	&	\textbf{2.7058	(0.2472)}	1.1684	(0.0654)	&	0.4539	(0.3896)	0.4962	(0.1601) \\
&	&	BETAASYM	&	3.2081	(0.9008)	1.2614	(0.1436)	&	0.6277	(0.7738)	0.5286	(0.0674) \\
&	&	KUM	&	3.2081	(0.9008)	1.2614	(0.1436)	&	0.6277	(0.7738)	0.5286	(0.0674) \\
&	&	TRI	&	2.7544	(0.2470)	 1.1852	(0.1281)	&	0.4581	(0.3436)	0.4858	(0.0333) \\
&	&	SN	&	5.7313	(0.3596)	1.7468	(0.0733)	&	0.5335	(0.0460)	0.5402	(0.0259) \\ \hline
												
&2048	&	CV	&	1.2935	(0.0816)	0.7886	(0.0289)	&	0.2480	(0.0161)	0.3332	(0.0128) \\
&	&	SURE	&	1.3517	(0.1128)	0.7959	(0.0337)	&	\textbf{0.2188	(0.0187)	0.3203	(0.0128)} \\
&	&	LPM	&	5.4387	(0.1660)	1.8619	(0.0312)	&	0.6011	(0.0210)	0.6183	(0.0112) \\
&	&	BETASYM	&	\textbf{1.1708	(0.2985)	0.7301	(0.0506)}	&	2.3630	(1.1634)	1.2162	(0.7111) \\
&	&	BETAASYM	&	1.3429	(0.4893)	0.7750	(0.1165)	&	2.3048	(1.1327)	1.1928	(0.6465) \\
&	&	KUM	&	1.3429	(0.4893)	0.7750	(0.1165)	&	2.3048	(1.1327)	1.1928	(0.6465) \\
&	&	TRI	&	1.4065	(0.3371)	0.8357	(0.0694)	&	2.5197	(1.1951)	1.3427	(0.7700) \\
&	&	SN	& 7.9610 (0.6565) 1.9464 (0.0410)	& 0.2607 (0.0149) 0.3512 (0.0128) \\ \hline

\end{tabular}
\caption{AMSEs, AMAEs and their standard deviations (SD) of the shrinkage and thresholding rules considered in the simulation study 2 for Donoho-Johnstone Bumps and Blocks test functions.}\label{tab:sim3}
\end{table}

\begin{table}[H]
\centering
\label{my-label}
\begin{tabular}{|c|c|c|c|c|}
\hline
Signal & $\boldsymbol{n}$ & \textbf{Method} & $\boldsymbol{\mathrm{SNR = 3}}$ & $\boldsymbol{\mathrm{SNR = 9}}$ \\ \hline \hline
&& & \textbf{AMSE (SD) AMAE (SD)} & \textbf{AMSE (SD) AMAE (SD)} \\ \hline \hline

Doppler&512	&	CV	&	1.2883	(0.1895)	0.8207	(0.0679)	&	0.3736	(0.0472)	0.4385	(0.0349)	\\
&	&	SURE	&	1.3336	(0.2111)	0.8367	(0.0700)	&	0.2164	(0.0304)	0.3367	(0.0268)	\\
&	&	LPM	&	5.4806	(0.3491)	1.8668	(0.0648)	&	0.6025	(0.0384)	0.6193	(0.0204)	\\
&	&	BETASYM	&	1.1316	(0.1769)	0.7838	(0.0594)	&	\textbf{0.1700	(0.0232)	0.2936	(0.0255)}	\\
&	&	BETAASYM	&	1.1867	(0.1947)	0.7928	(0.0610)	&	0.2597	(0.5393)	0.3313	(0.1598)	\\
&	&	KUM	&	1.1867	(0.1947)	0.7928	(0.0610)	&	0.2597	(0.0539)	0.3313	(0.1598)	\\
&	&	TRI	&	\textbf{1.1070	(0.1776)	0.7645	(0.0595)}	&	0.2799	(0.0305)	0.4245	(0.0293)	\\
&	&	SN	&	12.492 (0.2602) 	2.3993	(0.0571)	&	0.2953	(0.0327)	0.3986	(0.0252)	\\ \hline
													
&2048	&	CV	&	0.5560	(0.0508)	0.5207	(0.0286)	&	0.0973	(0.0089)	0.2118	(0.0122)	\\
&	&	SURE	&	0.5726	(0.0558)	0.5281	(0.0313)	&	\textbf{0.0966	(0.0088)	0.2127	(0.0120)}	\\
&	&	LPM	&	5.4426	(0.1847)	1.8618	(0.0333)	&	0.6035	(0.0193)	0.6196	(0.0105)	\\
&	&	BETASYM	&	0.4681	(0.6452)	0.4579	(0.1354)	&	0.4329	(1.6865)	0.3366	(0.3227)	\\
&	&	BETAASYM	&	0.4966	(0.7907)	0.4558	(0.1363)	&	0.2326	(0.2532)	0.2883	(0.0593)	\\
&	&	KUM	&	0.4966	(0.7907)	0.4558	(0.1363)	&	0.2326	(0.2532)	0.2883	(0.0593)	\\
&	&	TRI	&	\textbf{0.4350	(0.4131)	0.4424	(0.1275)}	&	0.1834	(0.6975)	0.2603	(0.1068)	\\
&	&	SN	&	4.3030 (0.1504) 1.2473 (0.0554)	&	0.1106 (0.0120) 0.2329 (0.0141)	\\ \hline \hline

Heavisine&512	&	CV	&	\textbf{0.5153	(0.0851)	0.5101	(0.0555)}	&	\textbf{0.1225	(0.0187)	0.2470	(0.0225)}	\\
&	&	SURE	&	0.5762	(0.0647)	0.5345	(0.0479)	&	0.2504	(0.0309)	0.3421	(0.0241)	\\
&	&	LPM	&	5.4480	(0.3349) 	1.8630	(0.0653)	&	0.6030	(0.0406)	0.6201	(0.0214)	\\
&	&	BETASYM	&	0.5931	(0.1184)	0.5440	(0.0639)	&	0.1331	(0.1057)	0.2586	(0.0629)	\\
&	&	BETAASYM	&	0.5376	(0.1003)	0.5101	(0.0587)	&	0.1780	(0.6275)	0.2722	(0.1328)	\\
&	&	KUM	&	0.5376	(0.1003)	0.5101	(0.0587)	&	0.1780	(0.6275)	0.2722	(0.1328)	\\
&	&	TRI	&	0.5384	(0.1077)	0.5098	(0.0580)	&	0.1931	(0.0225)	0.3187	(0.0209)	\\
&	&	SN	&	18.722 (0.9207)	3.6668	(0.1086)	&	0.2248	(0.0297)	0.3774	(0.0289)	\\ \hline
													
&2048	&	CV	&	\textbf{0.2651	(0.0307)	0.3494	(0.0235)}	&	\textbf{0.0608	(0.0067)	0.1634	(0.0115)}	\\
&	&	SURE	&	0.3656	(0.0344)	0.3964	(0.0249)	&	0.1115	(0.0327)	0.2158	(0.0350)	\\
&	&	LPM	&	5.4535	(0.1613)	1.8631	(0.0297)	&	0.6043	(0.0198)	0.6201	(0.0107)	\\
&	&	BETASYM	&	0.6101	(0.6674)	0.5377	(0.2205)	&	0.4579	(1.0737)	0.3718	(0.3347)	\\
&	&	BETAASYM	&	0.3156	(0.1443)	0.3905	(0.0963)	&	0.0969	(0.0334)	0.2384	(0.0338)	\\
&	&	KUM	&	0.3156	(0.1443)	0.3905	(0.0963)	&	0.0969	(0.0334)	0.2384	(0.0338)	\\
&	&	TRI	&	0.2713	(0.2367)	0.3473	(0.1008)	&	0.0859	(0.0121)	0.2276	(0.0188)	\\
&	&	SN	& 0.7543 (0.0710) 0.6561 (0.0365)	& 0.0687 (0.0076) 0.1798 (0.0134)	\\ \hline

\end{tabular}
\caption{AMSEs, AMAEs and their standard deviations (SD) of the shrinkage and thresholding rules considered in the simulation study 2 for Donoho-Johnstone Doppler and Heavisine test functions.}\label{tab:sim4}
\end{table}

\section{Illustration - IBOVESPA dataset}

\begin{figure}[H]
\centering
\subfigure[IBOVESPA dataset.]{
\includegraphics[scale=0.4]{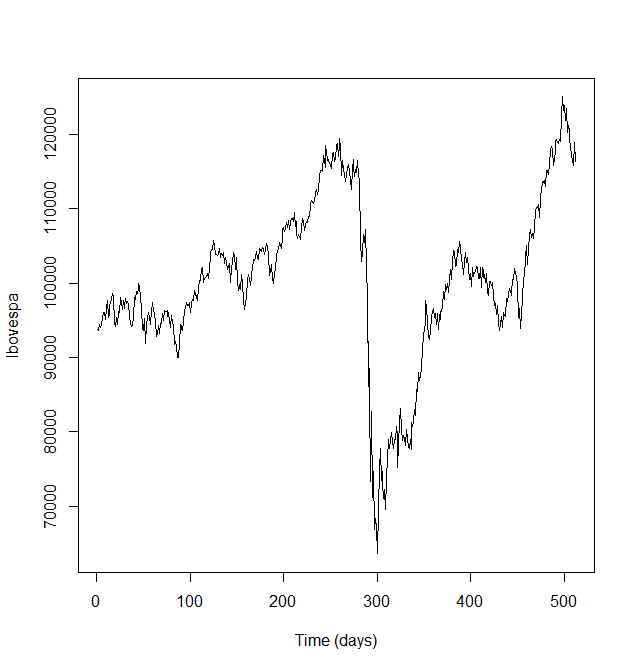}}
\subfigure[Empirical coefficients.]{
\includegraphics[scale=0.4]{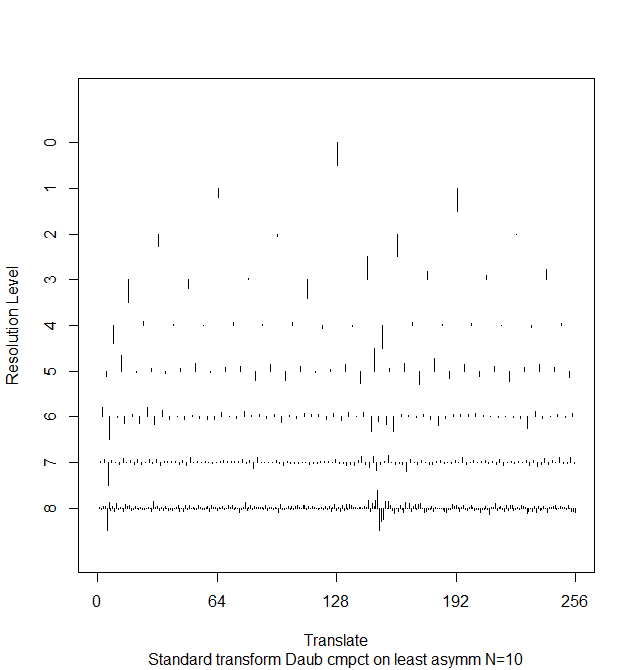}}
\caption{São Paulo stock market index time series(IBOVESPA, in Portuguese) from January 10, 2019 to January 29, 2021 (a) and its empirical wavelet coefficients after the application of a DWT with Daubechies basis with ten null moments (b).} \label{fig:data}
\end{figure}

Here we consider application of the proposed shrinkage rules to smooth the returns of the São Paulo stock market index (IBOVESPA, in Portuguese) time series from January 10, 2019 to January 29, 2021 ($n = 512$). This index is the most important in Brazil. Figure \ref{fig:data}(a) shows the time series. It is possible to note the occurrence of a big drop of the index around day 300, which corresponds to the beginnig of the COVID 19 pandemic. This feature can also be observed in the empirical wavelet coefficients of the dataset, after the application of a DWT with Daubechies basis with ten null moments, as depicted in Figure \ref{fig:data}(b). The significant empirical coefficients are located in the positions around the drop of the index and must be preserved in the shrinkage process.

\begin{figure}[H]
\centering
\subfigure[IBOVESPA - smooth version.]{
\includegraphics[scale=0.4]{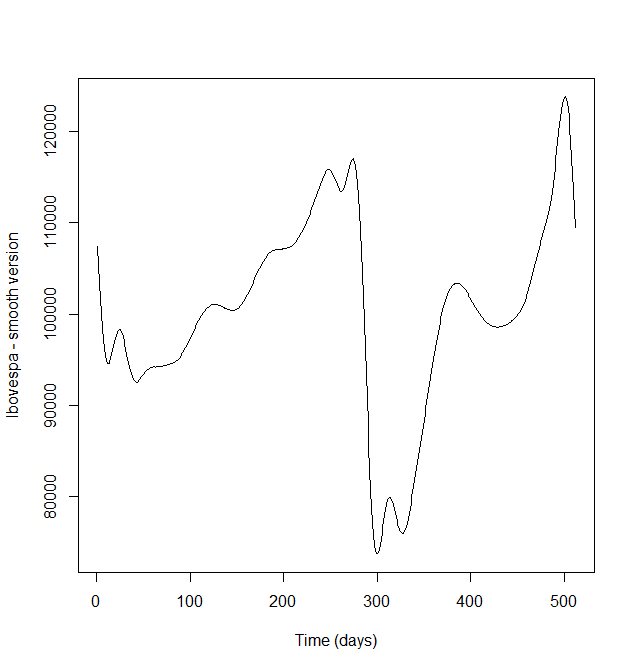}}
\subfigure[Shrunk (estimated) coefficients.]{
\includegraphics[scale=0.4]{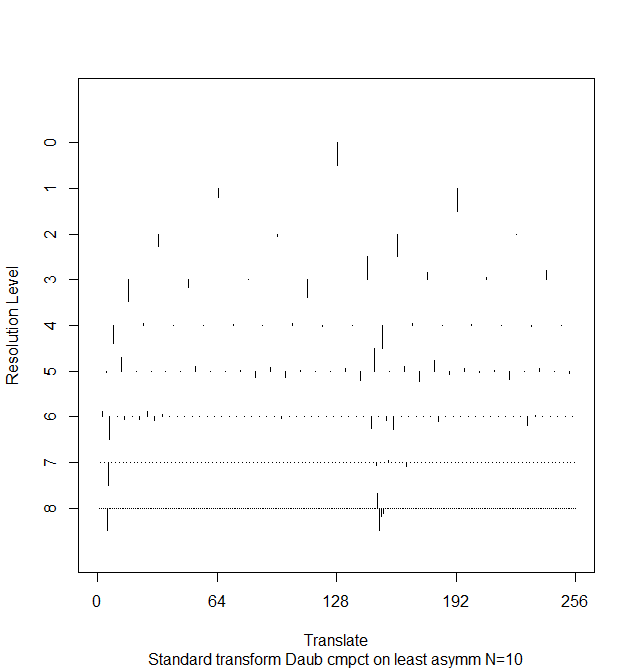}}
\caption{Smoothed version of the IBOVESPA time series after  application of the shrinkage rule under an asymmetric beta prior with $a = 5$ and $b=1$ (a) and the estimated wavelet coefficients (b).} \label{fig:data2}
\end{figure}

In order to smooth the time series and to obtain its main characteristics, we applied the shrinkage rules under asymmetric priors on the empirical coefficients. Since the smoothed versions of the time series were similar for all the shrinkage rules, Figure \ref{fig:data2}(a) shows the smoothed version of the data obtained after the application of the shrinkage rule under an asymmetric beta prior with hyperparameters $a = 5$ and $b=1$. Indeed, the denoised version clearly shows the main features of the index in terms of increasing and decreasing behaviors. Furthermore, $\hat{\sigma} = 875.64$ according to \eqref{eq:sigma} and the estimated $\mathrm{SNR}$ is 12.17, which can be considered a large signal-to-noise ratio, and consequently, a good value for the estimation process. Figure \ref{fig:data2}(b) presents the shrunk (estimated) wavelet coefficients associeted with the smoothed version of the dataset. Note that most of the empirical coefficients were shrunk by the rule, but the significant coefficients were preserved in magnitude.

\section{Final considerations}
This paper proposed a class of asymmetric priors to perform wavelet shrinkage in nonparametric regression models with additive Gaussian noise. The class is composed by the asymmetric beta, Kumaraswamy, asymmetric triangular and skew normal distributions and the priors are mixtures of a point mass function at zero and those densities. The idea is to provide a class of asymmetric priors to the wavelet coefficients that are flexible in shape and support. Among the proposed priors, the Kumaraswamy and the skew normal are novelties in terms of wavelet coefficient modeling but the triangular distribution has been considered in the literature only in the symmetric case.

The Bayesian shrinkage rules associated with the proposed priors are flexible in terms of severity of shrinking empirical coefficients and also allow the incorporation of their support in whether or not they are bounded by a symmetric interval (support in the entire real set). Furthermore, the hyperparameters are closely related to the degree of shrinkage of the rule, which facilitates their elicitations. 

The simulation studies indicated outstanding performance of the rules in estimating asymmetric wavelet coefficients versus some standard techniques, which was expected since these standard methods were obtained under the assumption of symmetric coefficients. In fact, the proposed rules overcame the standard methods in all scenarios of asymmetry, signal-to-noise ratio and sample size. The rules also had good performance in the simulations involving the Donoho-Johnstone test functions. In this simulation study, the rules performed better than the standard ones in some scenarios, mainly under the Bumps and Doppler underlying functions but also worked well under the Blocks and Heavisine functions despite the better performances of the standard techniques.

The impact of the chosen wavelet basis on the asymmetric shrinkage process and the behavior of the rules under non-Gaussian noise are suggestions of future works.

\end{document}